\documentclass[letterpaper, 12pt]{article}[2000/05/19]

\usepackage[english]{babel}

\usepackage{amsfonts,amsmath,amssymb,amsthm,latexsym,amscd,mathrsfs}
\usepackage{ifthen,cite}
\usepackage[bookmarksnumbered=true]{hyperref}
\hypersetup{pdfpagetransition={Split}}

\newcommand{\evenhead}{Author \ name}
\newcommand{\oddhead}{Article \ name}
\newcommand{\theArticleName}{Article name}

\newcommand{\FirstPageHeading}[1]{\thispagestyle{empty}%
\noindent\raisebox{0pt}[0pt][0pt]{\makebox[\textwidth]{\protect\footnotesize \sf }}\par}

\newcommand{\ArticleName}[1]{\renewcommand{\theArticleName}{#1}\vspace{-2mm}\par\noindent {\LARGE\bf  #1\par}}
\newcommand{\Author}[1]{\vspace{5mm}\par\noindent {\it #1} \par\vspace{2mm}\par}
\newcommand{\Address}[1]{\vspace{2mm}\par\noindent {\it #1} \par}
\newcommand{\Email}[1]{\ifthenelse{\equal{#1}{}}{}{\par\noindent {\rm E-mail: }{\it  #1} \par}}
\newcommand{\URLaddress}[1]{\ifthenelse{\equal{#1}{}}{}{\par\noindent {\rm URL: }{\tt  #1} \par}}
\newcommand{\EmailD}[1]{\ifthenelse{\equal{#1}{}}{}{\par\noindent {$\phantom{\dag}$~\rm E-mail: }{\it  #1} \par}}
\newcommand{\URLaddressD}[1]{\ifthenelse{\equal{#1}{}}{}{\par\noindent {$\phantom{\dag}$~\rm URL: }{\tt  #1} \par}}

\newcommand{\Keywords}[1]{\vspace{3mm}\par\noindent\hspace*{10mm}
\parbox{140mm}{\small {\it Key words:} \rm #1}\par}

\newcommand{\ShortArticleName}[1]{\renewcommand{\oddhead}{#1}}
\newcommand{\AuthorNameForHeading}[1]{\renewcommand{\evenhead}{#1}}

\long\def\@makecaption#1#2{%
  \vskip\abovecaptionskip
  \sbox\@tempboxa{\small \textbf{#1.}\ \ #2}%
  \ifdim \wd\@tempboxa >\hsize
    {\small \textbf{#1.}\ \ #2}\par
  \else
    \global \@minipagefalse
    \hb@xt@\hsize{\hfil\box\@tempboxa\hfil}%
  \fi
  \vskip\belowcaptionskip}

\def\E^#1{{\buildrel #1 \over\vee}}

\def\numberwithin#1#2{\@ifundefined{c@#1}{\@nocounterr{#1}}{%
  \@ifundefined{c@#2}{\@nocnterr{#2}}{%
  \@addtoreset{#1}{#2}%
  \toks@\@xp\@xp\@xp{\csname the#1\endcsname}%
  \@xp\xdef\csname the#1\endcsname
    {\@xp\@nx\csname the#2\endcsname
     .\the\toks@}}}}

\newtheorem{theorem}{Theorem}

{\theoremstyle{definition}

}

\begin {document}

\FirstPageHeading{Shtyk}

\ShortArticleName{The Existence and Uniqueness Theorem}

\ArticleName{The Existence and Uniqueness Theorem \\ for Solution of Generalized Hydrodynamical Equation}

\Author{V~O Shtyk}

\AuthorNameForHeading{V~O Shtyk}

\Address{ Bogolyubov Institute for Theoretical Physics of NAS of Ukraine}
%14-b, Metrolohichna str.  03680 Kyiv, Ukraine }

\EmailD{shtyk@bitp.kiev.ua}

%\Abstract%[Àíîòàö³ÿ]}
{\vspace{6mm}\par\noindent\hspace*{8mm}
\parbox{140mm}{\small { $\quad$

On the basis of
the sequence of marginal observables the evolution equations of the microscopic phase density and its generalizations is discussed.
We introduced dual BBGKY
hierarchy for these microscopic observables and their average values.

In the space of integrable functions, for initial-value problem of generalized hydrodynamical equation the existence and uniqueness theorem is proved.

 } }
}

%\Classification{37.05}
 \makeatletter
\renewcommand{\@evenhead}{
\hspace*{-3pt}\raisebox{-15pt}[\headheight][0pt]{\vbox{\hbox to \textwidth {\thepage \hfil \evenhead}\vskip4pt \hrule}}}
\renewcommand{\@oddhead}{
\hspace*{-3pt}\raisebox{-15pt}[\headheight][0pt]{\vbox{\hbox to \textwidth {\oddhead \hfil \thepage}\vskip4pt\hrule}}}
\renewcommand{\@evenfoot}{}
\renewcommand{\@oddfoot}{}
\makeatother
\newpage
\protect
\tableofcontents
\newpage

\section{Introduction}
The aim of the present paper is to derive the rigorous solution of the of generalized hydrodynamical equation for average value of marginal additive-type observable, known as marginal microscopic phase density. We proved the existence and uniqueness theorem for the such solution.

The paper is organized in the following order.

In section 2 we introduce the definitions used for the
description of the average values of observables and equivalent approach for the average values of marginal observables.
 In section 3 we give the definition of the microscopic phase density and its generalization. We consider the initial value problem for the dual BBGKY hierarchy for marginal observables and rewrite this hierarchy in terms of "macroscopic variables." Using the rule for average value in the grand canonical ensemble we derive the equations for average values of $k$-ary type of marginal microscopic phase density.
 In section 4 we consider the nonperturbative solution of initial value problem of such equations. This solution is presented over the particle clusters whose evolution is governed by the corresponding order cumulants of relevant evolution operators.
In section 5 we establish that in fact,if initial data is completely defined by a average value of marginal additive-type observable,then all possible average value of marginal $k$-type observable of infinitely many particle system at arbitrary moment of time
can be described within the framework of an average value of marginal additive-type observable by the hydrodynamical
equation without any approximations.
The existence and uniqueness theorem of a solution is presented in section 6.

\section{Preliminary facts}
We consider the system of a non-fixed
(i.e. arbitrary but finite) number of identical particles with unit mass $m=1$ in the space $\mathbb{R}^3$
(in the terminology of statistical mechanics it is known as \emph{nonequilibrium grand canonical ensemble}  \cite{CGP97}).
Every particle is characterized by the phase space coordinates  $x_i\equiv(q_i,p_i)$,
i.e. by the position on the space $q_i\in \mathbb{R}^3$ and momentum $p_i\in \mathbb{R}^3$.

An observable of this system can be described by the infinite sequence
$$A=(A_0,A_{1}(x_1),\ldots,A_{n}(x_1,\ldots,x_n),\ldots)$$ on $n$ particles.
The average values of observables (mean values or expectation values of observables) can written by the following formula
\begin{equation}\label{averageDA}
  \big\langle A\big\rangle(t)=\big(1,D(0)\big)^{-1}\big(A(t),D(0)\big)
  =\big(1,D(0)\big)^{-1}\sum\limits_{n=0}^{\infty}\frac{1}{n!}
  \int dx_{1}\ldots dx_{n}A_{n}(t)D_{n}(0)
\end{equation}
where
$\big(1,D(0)\big)={\sum\limits}_{n=0}^{\infty}\frac{1}{n!}
\int dx_{1}\ldots dx_{n}D_{n}(0)$ is a normalizing
factor (\emph{grand canonical partition function}).
The sequence $D(0)=(1,D_{1}(0,x_1),\ldots,D_{n}(0,x_1,\ldots,$ $x_n),\ldots)$
of probability densities of the distribution functions  $D_{n}(0)$ at the initial moment of time and the sequence of observables $A(t)=(A_0,A_{1}(t,x_1),\ldots,A_{n}(t,x_1,\ldots,x_n),\ldots)$
is a solution of initial-value problem of the Liouville equation for observables.
If $A(0)$ is the sequence of continuous functions and
$D(0)$ is the sequence of integrable functions, then functional (\ref{averageDA}) exists.

An equivalent approach of the description of evolution of many-particle systems,
that enables to describe systems in the thermodynamic
limit, is given by the sequences of  $s$-particle (marginal) observables
$G(t)=\big(G_0,G_{1}(t,x_1),\ldots,G_{s}(t,x_1,\ldots,x_s),\ldots\big)$. The sequence $G(t)$ is a solution of the initial-value problem of the dual Bogolyubov chain of equations (dual BBGKY hierarchy).
In that case the average values of observables in moment of time $t\in \mathbb{R}$ are determined by the functional \cite{CGP97}
\begin{equation}\label{avmarSch}
   \big\langle A \big\rangle(t)=\big(G(t),F(0)\big)=
   \sum\limits_{s=0}^{\infty}\frac{1}{s!}\int dx_{1}\ldots dx_s G_{s}(t)F_{s}(0),
\end{equation}
where the sequence of marginal observables $G(t)$ in terms of the sequence $A(t)$ are defined by the formula
\begin{equation}\label{mo}
   G_{s}(t,x_1,\ldots,x_s)=
   \sum_{n=0}^s\,\frac{(-1)^n}{n!}\sum_{j_1\neq\ldots\neq j_{n}=1}^s
   A_{s-n}\big(t,Y\backslash \{x_{j_1},\ldots,x_{j_{n}}\}\big),
\end{equation}
where $Y\equiv(x_1,\ldots,x_s)$,\, $s\geq 1$,
and the sequence $F(0)$ of marginal distribution functions is defined in terms of the sequence $D(0)$ as
follows (nonequilibrium grand canonical ensemble)
\begin{equation}\label{F(D)}
     F_{s}(0,x_1,\ldots,x_s)=
\big(1,D(0)\big)^{-1}
   \sum\limits_{n=0}^{\infty}\frac{1}{n!}\int dx_{s+1}\ldots dx_{s+n}D_{s+n}(0).
\end{equation}

\section{Evolution of marginal microscopic phase densities}

Let $N(t)\equiv\big(N^{(1)}(t),\ldots, N^{(k)}(t),\ldots\big)$, where
$N^{(k)}(t)=\big(0,\ldots,0,N_{k}^{(k)}(t),\ldots,N^{(k)}_{n}(t),\ldots\big)$,  $k\geq 1$,  be
the sequence of microscopic phase densities of $k$-ary type
\begin{equation}\label{k-arN}
    N_{n}^{(k)}(t)\equiv N_{n}^{(k)}(t,\xi_1,\ldots,\xi_k;x_1,\ldots,x_n)=
\sum\limits_{i_1\neq\ldots\neq i_k=1}^{n}\prod\limits_{l=1}^{k}\delta(\xi_{l}-X_{i_l}(t,x_1,\ldots,x_n)),
\end{equation}
where $\delta$ is the Dirac $\delta$-function, $\xi_1,\ldots,\xi_k$ are the macroscopic variables
$\xi_i=(v_i,r_i)\in\mathbb{R}^3\times\mathbb{R}^3$. Functions
$\big\{X_{i}(t,x_1,\ldots,x_n)\big\}_{i=1}^{n}$, $n\geq k\geq 1$, are the solution
of the Cauchy problem of the Hamilton equations for $n$ particles with the Hamiltonian
\begin{equation}\label{H}
  H_n=\sum\limits_{i=1}^{n}\frac{p_{i}^{2}}{2}+
  \sum\limits_{i<j=1}^{n} \Phi(q_{i}-q_{j}),
\end{equation}
where $\Phi(q_{i}-q_{j})$ is the two interaction potential, and with the initial data $x_1,\ldots,x_n$.

Thus, according to (\ref{mo}) at initial moment of time $t=0$  the sequence of marginal observables
of $k$-ary type microscopic phase densities (\ref{k-arN}) is given as follows
\begin{equation}\label{ad_G0}
   G^{(k)}(0)=\big(0,\ldots,0,\sum\limits_{i_1\neq\ldots\neq i_k=1}^{k}\prod\limits_{l=1}^{k}\delta(\xi_l-x_{i_l}),0,\ldots\big).
\end{equation}

We introduce the sequence of marginal observables \cite{GShZ}
$G(t)\equiv\big(G^{(1)}(t),\ldots,$ $ G^{(k)}(t),\ldots\big)$  of $k$-ary type
$G^{(k)}(t)=\big(0,\ldots,0,G_{k}^{(k)}(t),\ldots,G^{(k)}_{s}(t),\ldots\big).$

The marginal microscopic phase densities
$G^{(k)}_{s}(t)\equiv G^{(k)}_{s}(t,\xi_1,\ldots,\xi_k;x_1,\ldots,x_s)$  of every $k$-ary type
are governed by the initial-value problem of the dual BBGKY hierarchy \cite{BG}
\[
    \frac{\partial}{\partial t}G^{(k)}_{s}(t)=
    \big(\sum\limits_{i=1}^{s}\langle\, p_i,\frac{\partial}{\partial q_i}\rangle-
\sum\limits_{i\neq j=1}^{s}\langle\frac{\partial}{\partial q_i}\Phi(q_i-q_j),
      \frac{\partial}{\partial p_i}\rangle\big) G^{(k)}_{s}(t)-
\]
\begin{equation} \label{dual1}
      -\sum\limits_{i\neq j=1}^{s}\langle\frac{\partial}{\partial q_i}\Phi(q_i-q_j),
      \frac{\partial}{\partial p_i}\rangle G^{(k)}_{s-1}(t,Y\setminus x_{j})
\end{equation}
with the initial data
\begin{equation} \label{dual2}
         G^{(k)}_{s}(t)\mid_{t=0}=G^{(k)}_{s}(0),\quad s\geq k\geq 1.
\end{equation}
Here $Y\equiv(x_1,\ldots,x_s)$ and
$(x_1,\ldots,\E^{j},\ldots,x_s)\equiv(x_1,\ldots,x_{j-1},x_{j+1},\ldots,x_s)= Y\setminus x_{j}$.

Taking into account definition (\ref{ad_G0}) initial value problem (\ref{dual1})-(\ref{dual2}) can be rewritten in terms of variables $\xi.$
Indeed, in terms of variables $\xi_1,\ldots,\xi_k$ for the marginal microscopic phase densities of
$k$-ary type $G^{(k)}(t)=\big(0,\ldots,0,G_{k}^{(k)}(t),\ldots,G^{(k)}_{s}(t),\ldots\big)$
we derive
\[
  \frac{\partial}{\partial t}G^{(k)}_{s}(t)=\big(-\sum\limits_{i=1}^{k}\langle v_i,\frac{\partial}{\partial r_i}\rangle+
\sum\limits_{i\neq j=1}^{k}\langle \frac{\partial}{\partial r_i}\Phi(r_i-r_j),
   \frac{\partial}{\partial v_i}\rangle \big)G^{(k)}_{s}(t)+
\]
\begin{equation} \label{k_lanc}
     +\sum\limits_{i=1}^{k}\int d\xi_{k+1}\langle\frac{\partial}{\partial r_i}\Phi(r_i-r_{k+1}),
     \frac{\partial}{\partial v_i}\rangle G^{(k+1)}_{s}(t)
\end{equation}
with the initial data
\begin{equation} \label{k_lancin}
    G^{(k)}_{s}(t)\mid_{t=0}=\sum\limits_{i_1\neq\ldots\neq i_k=1}^{s}
    \prod\limits_{l=1}^{k}\delta(\xi_l-x_{i_l})\delta_{s,k},
\end{equation}
where $1\leq r<s$ and for $k=s$, the marginal microscopic phase density $G^{(s)}_{s}(t)$ is governed by the Liouville equation.

It should be noted that in terms of variables $\xi_1,\ldots,\xi_k$
dual BBGKY hierarchy (\ref{k_lanc})
is represented as the Bogolyubov set of equations with respect to
the arity index $k\geq1$,
while evolution equations (\ref{k_lanc}) have a structure of a sequence of equations
with respect to the index of number of particles $s\geq k$.

Averaging initial value problem (\ref{k_lanc})-(\ref{k_lancin}) the evolution equation of average value (\ref{avmarSch})
 for $k$-ary type microscopic
phase density, according to (\ref{avmarSch}) has the form \cite{GShZ}
\[
    \frac{\partial}{\partial t}\langle G^{(k)} \rangle (t)=-\sum\limits_{i=1}^{k}\big\langle v_i,
    \frac{\partial}{\partial r_i}\big\rangle \langle G^{(k)} \rangle (t)+
\sum\limits_{i\neq j=1}^{k} \langle \frac{\partial}{\partial r_i}\Phi(r_i-r_j),
    \frac{\partial}{\partial v_i} \rangle \langle G^{(k)} \rangle (t)+
\]
\begin{equation}\label{Gs1}
    +\sum\limits_{i=1}^{k}\int d\xi_{k+1} \langle\frac{\partial}{\partial r_i}\Phi(r_i-r_{k+1}),
    \frac{\partial}{\partial v_i} \rangle \langle G^{(k+1)} \rangle (t),
\end{equation}
with the initial data
\begin{equation}\label{Gs2}
    \langle G^{(k)} \rangle (t,\xi_1,\ldots,\xi_k)|_{t=0}=\langle G^{(k)} \rangle(0), \quad k\geq1.
\end{equation}
We remark that due to functional (\ref{avmarSch}) initial data (\ref{Gs2}) are given as the following functions
\[
   \langle G^{(k)} \rangle(0,\xi_1,\ldots,\xi_k)=F_{k}(0,\xi_1,\ldots,\xi_k),
\]
where
$F_s(0,\xi_1,\ldots,\xi_s)$ is the value of the solution of BBGKY hierarchy at initial moment of time at a point $\xi_1,\ldots,\xi_s$.

\section{On solution of initial-value problem for the average value of marginal microscopic phase density}

To derive the hydrodynamic equation for one-particle function it is necessary to construct a non-perturbative solution of Cauchy problem (\ref{Gs1})-(\ref{Gs2}) in the following form.

Consider a solution of Cauchy problem (\ref{Gs1})-(\ref{Gs2}) that
is defined by the expansion
over the arity index of the microscopic phase density whose evolution are governed by
the corresponding-order cumulant (semi-invariant) of the
evolution operators (\ref{eo}), constructed in \cite{GShZ}
\[
\langle G^{(k)} \rangle (t,\xi_1,\ldots,\xi_k)=\big( U(t)\langle G \rangle (0)\big)_k(\xi_1,\ldots,\xi_k)=
\]
\begin{equation}\label{RozvG}
        =\sum\limits_{n=0}^{\infty}\frac{1}{n!}\int d\xi_{k+1}\ldots d\xi_{k+n}
      \mathfrak{A}_{1+n}(-t)\langle G^{(k+n)}\rangle (0),
\end{equation}
where $G(0)=(0,G^{(1)}(0),\ldots,G^{(k)}(0),\ldots)$ is the sequence on integrable functions and if $n \geq 0$ the operator
\begin{equation}\label{cm}
\mathfrak{A}_{1+n}(-t)\equiv\mathfrak{A}_{1+n}(-t,\{Y\},{k+1},\ldots,{k+n})=
\end{equation}
\[
=\sum\limits_{\mathrm{P}:\{\{Y\},X\setminus Y\} ={\bigcup\limits}_i X_i}(-1)^{|\mathrm{P}|-1}(|\mathrm{P}|-1)!
        \prod_{X_i\subset \mathrm{P}}S_{|X_i|}(-t,X_i)
\]
is the $(1+n)th$-order cumulant of the groups of operators (\ref{eo}), ${\sum\limits}_\mathrm{P}$
is the sum over all possible partitions $\mathrm{P}$ of the set $\{\{Y\},{k+1},\ldots,{k+n}\}$ into
$|\mathrm{P}|$ nonempty mutually disjoint subsets $X_i\subset \{\{Y\},X\setminus Y\}\equiv\{\{Y\},{k+1},\ldots,{k+n}\}$.
The set $\{Y\}$ consists of one element of $Y\equiv({1},\ldots,{k})$,
i.e. the set $({1},\ldots,{k})$ is a connected subset of the partition $\mathrm{P}$ ($|\mathrm{P}|=1$).

If $\langle G^{(k)}\rangle (0)$ are integrable functions \cite{GerRS}, series (\ref{RozvG}) converges for small densities.

In (\ref{cm}) evolution operators $S_{k}(-t)$ act as follows
\begin{equation}\label{eo}
\big(S_{k}(-t)f_{k}\big)(\xi_1,\ldots,\xi_k):=f_{k}(\Xi_1(-t,\xi_1,\ldots,\xi_k),\ldots,\Xi_k(-t,\xi_1,\ldots,\xi_k)),
\end{equation}
where functions $\Xi_i(t)\equiv\big(V_i(t),R_i(t)\big)$  are the solution of the Cauchy problem of
the Hamilton equations for "macroscopic variables"\,
\[
\frac{d }{d t}R_i(t)=V_i(t),\qquad
\frac{d}{d t} V_i(t)=-\sum\limits_{j\neq i, j=1}^{k}\frac{\partial}{\partial R_{i}(t)}\Phi\big(R_i(t)-R_j(t)\big),
\]
with the initial data
\[
R_i(0)=r_i,\qquad V_i(0)=v_i,\qquad i=1,\ldots, k.
\]

We remark that using Duhamel formula solution (\ref{RozvG}) of Cauchy problem (\ref{Gs1})-(\ref{Gs2}) can be represented as the iteration series
of BBGKY hierarchy (\ref{Gs1}) \cite{GShZ}.

%%%%%%%%%%%%%%%%%%%%%%%%%%%%%%%%%%%%%%%%%%%%%%%%%%%%%%%%%%%%%%%%%%%%%%%%%%%%%%%%%%%%%%%%%%%%%%%%%%%%%%%%%%%%%%%%%%%%%%%%%%%%%%%%%%%%%%%%%%%%%%%%%%%

\section{Hydrodynamical equations for average values of observables}
Consider initial data that completely defined by one-particle function $\langle G^{(1)} \rangle (0,\xi_1)$, i.e.
 $$\langle G \rangle (0)=\big(I,\langle G^{(1)} \rangle (t,\xi_1),\ldots,\prod_{i=1}^k \langle G^{(1)} \rangle (t,\xi_i),\ldots\big).$$ Taking into account that  $\langle G^{(1)}\rangle(0,\xi_i)=F_{1}(0,\xi_i),$  initial data  satisfy the chaos condition.

For such initial data there is a possibility to reformulate initial value problem (\ref{Gs1})-(\ref{Gs2}) as new one for unknown function  $ \langle G^{(1)} \rangle (t,\xi_1)$ with the functional $\langle G^{(k)} \rangle \big(t,\xi_1,\ldots,\xi_k \mid \langle G^{(1)} \rangle (t)\big),\, k\geq 2 $.

Functionals $\langle G^{(k)} \rangle \big(t,\xi_1,\ldots,\xi_k \mid \langle G^{(1)} \rangle (t)\big),\, k\geq 2 $ are represented by the following expansion over the products of one-particle functions  $\langle G^{(1)} \rangle(t)$
\begin{multline}\label{f}
   \langle G^{(k)} \rangle\big(t,\xi_1,\ldots,\xi_k \mid \langle G^{(1)} \rangle(t)\big):=\\
   =\sum\limits_{n=0}^{\infty }\frac{1}{n!}\int d\xi_{k+1}\ldots d\xi_{k+n} \,\mathfrak{V}_{1+n}\big(t,\{Y\},k+1,\ldots,k+n\big)\prod _{i=1}^{k+n} \langle G^{(1)} \rangle(t,\xi_i),
\end{multline}
where evolution operators $\mathfrak{V}_{1+n}(t),\, n\geq0$
defined as follows \cite{GG}
\begin{eqnarray}\label{kk}
        &&\mathfrak{V}_{1+n}(t,\{Y\},X \backslash Y):= n! \sum_{r=0}^{n}(-1)^r \sum_{m_1=1}^{n}\ldots\sum_{m_r=1}^{n-m_1-\ldots-m_{r-1}}\frac{1}{(n-m_1-\ldots-m_r)!}\times\nonumber\\\
           &&\times\widehat{\mathfrak{A}}_{1+n-m_1-\ldots-m_r}(t,\{Y\},k+1,\ldots,k+n-m_1-\ldots-m_r)\times\nonumber\\
        &&\times\prod_{j=1}^{r}\sum_{r_2^j=0}^{m_j}\ldots\sum_{r_{n-m_1-\ldots-m_j+k}^j=0}^{r_{n-m_1-\ldots-m_j+k-1}^j}
        \prod_{i_j=1}^{k+n-m_1-\ldots-m_j}
           \frac{1}{(r_{n-m_1-\ldots-m_j+k+1-i_j}^j-r_{n-m_1-\ldots-m_j+k+2-i_j}^j)!} \times\nonumber\\
           &&\times
           \widehat{\mathfrak{A}}_{1+r_{n-m_1-\ldots-m_j+k+1-i_j}^j-r_{n-m_1-\ldots-m_j+k+2-i_j}^j}(t,i_j,k+n-m_1-\ldots-m_j+1+\nonumber\\
          && +r_{k+n-m_1-\ldots-m_j+2-i_j}^j,\ldots,k+n-m_1-\ldots-m_j+r_{k+n-m_1-\ldots-m_j+1-i_j}^j),
\end{eqnarray}
where $r_1^j=m_j,$ $r_{n-m_1-\ldots-m_j+k+1}^j=0.$
We denoted
$\widehat{\mathfrak{A}}_{1+n}(t)$ is the $(n+1)$-th order cumulant
 of scattering operators $ \widehat{S}_{k}(t),\,k\geq1 \,$
($\widehat{S}_{1}(t)=I$ is a unit operator)
\begin{eqnarray}\label{so}
     \widehat{S}_{k}(t,1,\ldots,k)=S_{k}(-t,1,\ldots,k)
     \prod _{i=1}^{k}S_{1}(t,i).
\end{eqnarray}
The samples of two first terms of evolution operator $\mathfrak{V}(t)$ are the following
\begin{eqnarray*}
   &&\mathfrak{V}_{1}(t,\{Y\})=\widehat{\mathfrak{A}}_{1}(t,\{Y\}),\\
   &&\mathfrak{V}_{2}(t,\{Y\},k+1)=\widehat{\mathfrak{A}}_{2}(t,\{Y\},k+1)-\widehat{\mathfrak{A}}_{1}(t,\{Y\})
          \sum_{i_1=1}^k \widehat{\mathfrak{A}}_{2}(t,i_1,k+1),
   %&&\mathfrak{V}_{3}(t,\{Y\},s+1,s+2)=\widehat{\mathfrak{A}}_{3}(t,\{Y\},s+1,s+2)-2!\widehat{\mathfrak{A}}_{2}(t,\{Y\},s+1)\times\\
%         &&\times \sum_{i_1=1}^{s+1} \widehat{\mathfrak{A}}_{2}(t,i_1,s+2)-\widehat{\mathfrak{A}}_{1}(t,\{Y\})\big(\sum_{i_1=1}^{s} \widehat{\mathfrak{A}}_{3}(t,i_1,s+1,s+2)-\\
%         &&-2!\sum_{i_1=1}^{s}\sum_{i_2=1}^{s+1}\widehat{\mathfrak{A}}_{2}(t,i_1,s+1)\widehat{\mathfrak{A}}_{2}(t,i_2,s+2)+
%         2!\sum_{1=i_1<i_2}^{s}\widehat{\mathfrak{A}}_{2}(t,i_1,s+1)\widehat{\mathfrak{A}}_{2}(t,i_2,s+2)\big),
\end{eqnarray*}
or in terms of scattering operators
\begin{eqnarray*}
    && \mathfrak{V}_{1}(t,\{Y\})=\widehat{S}_{k}(t,Y),\\
    &&\mathfrak{V}_{2}(t,\{Y\},k+1)= \widehat{S}_{k+1}(t,Y,k+1)-\widehat{S}_k(t,Y)
       \sum_{j=1}^s  \widehat{S}_2(t,j,k+1)+(k-1)\widehat{S}_k(t,Y).
\end{eqnarray*}

On integrable functions the action of scattering operators (\ref{so}) are defined by the following formula
\begin{equation}\label{sod}
 \big(\widehat{S}_{k}(t)f_{k}\big)(\xi_1,\ldots,\xi_k)=f_{k}\Big(\Xi_1\big(t,\,\Xi_1(-t,\xi_1,\ldots,\xi_k)\big),\ldots
,\Xi_k\big(t,\,\Xi_k(-t,\xi_1,\ldots,\xi_k)\big)\Big),
\end{equation}
where functions $\Xi_i(t)\equiv\big(V_i(t),R_i(t)\big)$,\,$i=1,\ldots,n$  are solutions of the Cauchy problem of
Hamilton equations for "macroscopic variables"\, with corresponding initial data.

The generator of first-order evolution operator (\ref{kk}), i.e. $n=0$, is defined by the Poisson bracket
with respect to the variables $\xi_1,\ldots,\xi_k$
with an interaction potential $\Phi$ on the continuously differentiable functions $f_{k}\equiv f_{k}(\xi_1,\ldots,\xi_k)$
\[
  \lim\limits_{t\rightarrow 0}\frac{1}{t}\big(\mathfrak{V}_{1}(t,\{Y\})-I\big) f_{k}=
  \sum\limits_{i\neq j=1}^{k}\langle \frac{\partial}{\partial r_i}\Phi(r_i-r_j),
  \frac{\partial}{\partial v_i}\rangle f_{k}.
\]

In generale case, for  $n\geq 1,$ it holds
\[
  \lim\limits_{t\rightarrow 0}\frac{1}{t}\big(\mathfrak{V}_{1+n}(t,\{Y\},X\setminus Y)\big) f_{k+n}=0.\]

For the case of non-interacting particles the evolution operators (\ref{kk}) for $n\geq 0$ we have
\[
  \lim\limits_{t\rightarrow 0}\frac{1}{t}\big(\mathfrak{V}_{1+n}(t,\{Y\},X\setminus Y)\big) f_{k+n}=\delta_{n,0}.\]
At the same time for initial moment of time $t=0$
\[
  \lim\limits_{t\rightarrow 0}\frac{1}{t}\big(\mathfrak{V}_{1+n}(t,\{Y\},X\setminus Y)\big) f_{k+n}=\delta_{n,0},\]
where $\delta_{n,0}$ is a Kronecker symbol.

The average value for an additive-type microscopic phase density $\langle G^{(1)}\rangle(t)$ is governed by
the following Cauchy problem (\emph{the generalized evolution
equation for the average value of microscopic phase density})
\begin{equation}\label{gke}
\frac{\partial}{\partial t}\langle G^{(1)}\rangle(t,\xi_1)+\langle v_1,\frac{\partial}{\partial r_1}\rangle\langle G^{(1)}\rangle(t,\xi_1)=
\end{equation}
\[
   =\int d\xi_{2}\langle\frac{\partial}{\partial r_1}\Phi(r_1-r_2),\frac{\partial}{\partial v_1}\rangle
\langle G^{(2)}\rangle\big(t,\xi_1,\xi_2 \mid \langle G^{(1)}\rangle(t)\big)
\]
with the initial data
\begin{equation}\label{gke2}
           \langle G^{(1)}\rangle(t,\xi_1)|_{t=0}= \langle G^{(1)}\rangle(0,\xi_1).
\end{equation}
where functional $\langle G^{(2)}\rangle\big(t,\xi_1,\xi_2 \mid \langle G^{(1)}\rangle(t)\big)$ is defined by  (\ref{f}).

The functional $\langle G^{(2)}\rangle\big(t,\xi_1,\xi_2 \mid \langle G^{(1)}\rangle(t)\big)$ in the collision integral
of equation (\ref{gke}) is defined by expansion (\ref{f})
\begin{equation}\label{ff}
\langle G^{(2)}\rangle(t)\big(t,\xi_1,\xi_2 \mid \langle G^{(1)}\rangle(t)\big)=
\end{equation}
\[
=\widehat{\mathfrak{A}}_{1}(t,\{Y\})\prod_{i=1}^2 \langle G^{(1)}\rangle(t,\xi_i)+
\int d\xi_{3}\big(\widehat{\mathfrak{A}}_{2}(t,\{Y\},\xi_{k+1})-
\]
\[
 - \widehat{\mathfrak{A}}_{1}(t,\{Y\})\sum_{j=1}^s \widehat{\mathfrak{A}}_{2}(t,\xi_{j},\xi_{k+1}) \big)
 \prod_{i=1}^{3} \langle G^{(1)}\rangle(t,\xi_i) + ...,
\]
where $\{Y\}\equiv(\xi_1,\xi_2)$.

We represent the first term of expansion (\ref{ff}) in an explicit form.
According to the definition of two-particle scattering operator (\ref{sod}) we have
\begin{equation}\label{ffff}
  \widehat{\mathfrak{A}}_{1}(t,\{Y\})\prod_{i=1}^2 \langle G^{(1)}\rangle(t,\xi_i)=
\end{equation}
\[
  =\langle G^{(1)}\rangle \Big(t,\,\Xi_1\big(t,\,\Xi_1(-t,\,\xi_1,\xi_2)\big)\Big)
\langle G^{(1)}\rangle \Big(t,\,\Xi_2\big(t,\,\Xi_2(-t,\,\xi_1,\xi_2)\big)\Big),
\]
where the functions ($i=1,2$)
\[
  \Xi_i\big(t,\,\Xi_i(-t,\,\xi_1,\xi_2)\big)=
\big(V_i(-t,\,\xi_1,\xi_2),R_i(-t,\,\xi_1,\xi_2)+t\,V_i(-t,\,\xi_1,\xi_2)\big)
\]
are defined as above in (\ref{sod}).

\section{The existence and uniqueness theorem}
Suppose $L^{1}_{n}$ is the Banach space of integrable
functions defined on the phase space $\mathbb{R}^{3n}\times\mathbb{R}^{3n}$, of $n$-particle
system, symmetric under the perturbations of arguments. The norm of an element $f_{n}$ of $L^{1}_{n}$ is denoted by
\begin{equation*}
    \|f_{n}\|=\int_{\mathbb{R}^{3}\times\mathbb{R}^{3}} dx_{1}\ldots dx_{n}
    |f_{n}(x_{1},\ldots,x_{n})|,
   \end{equation*}
   $L^{1}_{n,0}\subset L^{1}_n$ is a subspace of continuously
differentiable functions with compact supports.
For functionals of average values of observables (\ref{f}) the following existence theorem is true \cite{GG}

\begin{theorem}
If $||\langle G^{(1)}\rangle(0)||_{L^1_1(\mathbb{R}^3\times\mathbb{R}^3)}<e^{-10}(1+e^{-9})^{-1}$ then there are exist a global in time solution of the initial value problem (\ref{gke})-(\ref{gke2}) of the generalized kinetic equation which is defined by
\begin{equation}\label{Rozvgke}
\langle G^{(1)} \rangle (t,\xi_1) =\sum\limits_{n=0}^{\infty}\frac{1}{n!}\int d\xi_{2}\ldots d\xi_{1+n}\,
      \mathfrak{A}_{1+n}(-t)\prod_{i=1}^{n+1} \langle G^{(1)}\rangle(0,\xi_i),
\end{equation}
For $\langle G^{(1)}\rangle(0)\in L^1_{1,0}(\mathbb{R}^3\times\mathbb{R}^3)$  it is a strong solution and for arbitrary initial data it is a weak solution.
\end{theorem}

\begin{proof}
We give a short proof of the theorem here. The full version can be found in \cite{GG}.

Taking into account the fact that operator $S(-t)$ is the isometric operator and for $|X\setminus Y|>1$ the identity
\[\sum_{P:\{X\setminus Y\}=\cup_iX_i}(-1)^{|P|-1}(|P|-1)!=0\]
is valid, for cumulants $\mathfrak{A}(t)$ we have
\begin{equation}\label{RozvGn}
        \int\limits_{\mathbb{R}^3\times\mathbb{R}^3} d\xi_{k+1}\ldots d\xi_{k+n}
      \mathfrak{A}_{n}(t,k+1,\ldots,k+n)\prod\limits_{i=1}^{k+n}\langle G^{(1)}\rangle (t,\xi_i)=0,
\end{equation}
Therefore, we can represent marginal functionals of the average value of observables by the following renormalized series

\begin{multline}\label{frr}
   \langle G^{(k)} \rangle\big(t,\xi_1,\ldots,\xi_k \mid \langle G^{(1)} \rangle(t)\big):=\\
   =\sum\limits_{n=0}^{\infty }\frac{1}{n!}\int d\xi_{k+1}\ldots d\xi_{k+n} \,\mathfrak{\tilde{V}}_{1+n}\big(t,\{Y\},k+1,\ldots,k+n\big)\prod _{i=1}^{k+n} \langle G^{(1)} \rangle(t,\xi_i),
\end{multline}
where evolution operators $\mathfrak{\tilde{V}}_{1+n}(t),\, n\geq0$
defined as follows
\begin{eqnarray}\label{kkr}
        &&\mathfrak{\tilde{V}}_{1+n}(t,\{Y\},X \backslash Y):= n! \sum_{r=0}^{n}(-1)^r \sum_{m_1=1}^{n}\ldots\sum_{m_r=1}^{n-m_1-\ldots-m_{r-1}}\frac{1}{(n-m_1-\ldots-m_r)!}\times\nonumber\\\
           &&\times\widehat{\mathfrak{A}}_{1+n-m_1-\ldots-m_r}(t,\{Y\},k+1,\ldots,k+n-m_1-\ldots-m_r)\times\nonumber\\
        &&\times\prod_{j=1}^{r}\sum_{r_2^j=0}^{m_j}\ldots\sum_{r_{k}^j=0}^{k-1}
        \prod_{i_j=1}^{k}
           \frac{1}{(r_{k+1-i_j}^j-r_{k+2-i_j}^j)!} \times\nonumber\\
           &&\times
           \widehat{\mathfrak{A}}_{1+r_{k+1-i_j}^j-r_{k+2-i_j}^j}(t,i_j,k+n-m_1-\ldots-m_j+1+\nonumber\\
          && +r_{k+2-i_j}^j,\ldots,k+n-m_1-\ldots-m_j+r_{k+1-i_j}^j).
\end{eqnarray}
Then we use the following estimate for the cumulants $\mathfrak{A}(t)$
\begin{equation*}\label{RozvGn}
        \int\limits_{\mathbb{R}^3\times\mathbb{R}^3} d\xi_{1}\ldots d\xi_{k+n}
      \big|\mathfrak{A}_{1+n}(-t,\{Y\},k+1,\ldots,k+n)\prod\limits_{i=1}^{k+n}\langle G^{(1)}\rangle (t,\xi_i)\big|\leq n!e^{n+2}||\langle G^{(1)}\rangle(t)||^{k+n}_{L^1_1},
\end{equation*}
Using the last estimate for (\ref{frr}) we obtain
\begin{eqnarray}\label{1}
              ||\langle G^{(k)}\rangle(t|\langle G^{(1)}\rangle)||_{L^1_s}\leq ||\langle G^{(1)}\rangle(t)||^{k}_{L^1_1}e^{2}\sum\limits_{n=0}^{\infty}||\langle G^{(1)}\rangle(t)||^{n}_{L^1_1}e^{n}+\nonumber\\
              +||\langle G^{(1)}\rangle(t)||^{k}_{L^1_1}e^{3k+2}\sum\limits_{n=1}^{\infty}||\langle G^{(1)}\rangle(t)||^{n}_{L^1_1}e^{(3k+2)n},
\end{eqnarray}
Hence the sequence of marginal functionals of the state $\langle G^{(k)}\rangle(t,\xi_1,\ldots,\xi_k|\langle G^{(1)}\rangle),$ $k\geq2$ exists and represents by converged series (\ref{f}) provided that
\[||\langle G^{(1)}\rangle(t)||_{L^1_1}<e^{-(3k+2)}.\]
On the other hand, using the estimate for series (\ref{Rozvgke}), we have
\begin{eqnarray}\label{2}
              ||\langle G^{(1)}\rangle(t)||_{L^1_1}\leq ||\langle G^{(1)}\rangle(0)||_{L^1_1}e^{2}\sum\limits_{n=0}^{\infty}||\langle G^{(1)}\rangle(t)||^{n}_{L^1_1}e^{2}\sum\limits_{n=0}^{\infty}e^n||\langle G^{(1)}\rangle(0)||^{n}_{L^1_1}.
\end{eqnarray}
Finally, according to estimates (\ref{1}) and (\ref{2}), we conclude that the collision integral of generalized hydrodynamical equation exists under the condition on initial data
  $$||\langle G^{(1)}\rangle(0)||_{L^1_1(\mathbb{R}^3\times\mathbb{R}^3)}<e^{-10}(1+e^{-9})^{-1}.$$
\end{proof}

\section*{Acknowledgement}
This work was supported by the grant ¹GP/F32/077.


\addcontentsline{toc}{section}{References}

\begin{thebibliography}{99}

\bibitem{Ba95}  R. Balescu, Phys.~Rev., \textbf{E 51}, 4807, (1995).
\bibitem{Ba97}  R. Balescu, Phys.~Rev., \textbf{E 55}, 2465, (1997).
\bibitem{KL}    E.M. Kramer, A.E. Lobkovsky, Phys.~Rev., \textbf{E 53}, 1465, (1996).
\bibitem{A}     F. Amblard et al.,  Phys.~Rev. Lett., \textbf{77},  4470, (1996).
\bibitem{MK}    R. Metzler, J. Klafter, Phys.~Rep., \textbf{339}, 1, (2000).
\bibitem{SK}    I.M. Sokolov, J. Klafter, Chaos, \textbf{15}, 026103, (2005).
\bibitem{Ba05}  R. Balescu, \textit{Aspects of anomalous transport in plasmas.} (Bristol: Institute of Physics, 2005).
\bibitem{ZW99}  A. Zagorodny, J. Weiland, Phys. Plasmas, \textbf{6}, 2359, (1999).
\bibitem{ZW01}  A. Zagorodny, J. Weiland, Condens. Matter Phys., \textbf{4}, 603, (2001).
\bibitem{ZW09}  A. Zagorodny, J. Weiland, Phys. Plasmas, \textbf{16}, 052308, (2009).
\bibitem{OK}    O.Z. Orszag, R.H. Kraichnan, Phys. Fluids, \textbf{10}, 1720, (1967).
\bibitem{GHT}    H. Grabert, P. Hanggi, P. Talkner, J. Stat. Phys., \textbf{22}, 537, (1980).
\bibitem{Ba}    R. Balescu, \textit{Equilibrium and Nonequilibrium  Statistical Mechanics.} (N.Y.: John Wiley$\&$Sons, 1975).
\bibitem{BC}    N.N. Bogolyubov, \textit{Problems of Dynamic Theory in Statistical Physics.} (M.-L.: FITTL, 1946).
\bibitem{CGP97} C. Cercignani, V.I. Gerasimenko, D.Ya. Petrina, \textit{Many-Particle Dynamics and Kinetic Equations.} (Kluwer Acad. Publ., 1997).
\bibitem{GG}    I.V. Gapyak, V.I. Gerasimenko, On Rigorous Derivation of the Enskog Kinetic Equation, Preprint: ArXiv:1107.5572v1
\bibitem{GShZ}  V.I. Gerasimenko, V.O. Shtyk, A.G. Zagorodny, Ukr. J. Phys., \textbf{54}, 8-9, 795-807, (2009).
\bibitem{BG}    G. Borgioli, V.I. Gerasimenko, Riv. Mat. Univ. Parma., \textbf{4}, 251 (2001).
\bibitem{GerRS} V.I. Gerasimenko, T.V. Ryabukha, M.O. Stashenko, J. Phys. A: Math. Gen. \textbf{37}, 9861 (2004).
\bibitem{GP98}  V.I. Gerasimenko, D.Ya. Petrina, Ukrainian J. Phys., \textbf{43}, 697 (1998).
\bibitem{Co09}  E.G.D. Cohen, Math. Mod Meth. Appl. Sci., \textbf{7}, 909 (1997).
\bibitem{GP90}  V.I. Gerasimenko, D.Ya. Petrina, Uspekhi Mat. Nauk., \textbf{45}, 135 (1990).
\bibitem{Sh} Shtyk V~O 2007 On the solutions of the nonlinear Liouville hierarchy
                {\it J. Phys. A: Math. Theor.} {\bf 40} 9733-9742

\end{thebibliography}
\end{document}